\documentclass[twocolumn]{aastex631}

\usepackage [english]{babel}
\usepackage [autostyle, english = american]{csquotes}
\MakeOuterQuote{"}

\usepackage{amsmath}

\begin{document}

\title{Analytical models for secular descents in hierarchical triple systems}

\correspondingauthor{Grant C. Weldon}
\email{gweldon@astro.ucla.edu}

\author{Grant C. Weldon}
\affiliation{Department of Physics and Astronomy, UCLA, Los Angeles, CA 90095, USA}
\affiliation{Mani L. Bhaumik Institute for Theoretical Physics, Department of Physics and Astronomy, UCLA, Los Angeles, CA 90095, USA}

\author{Smadar Naoz}
\affiliation{Department of Physics and Astronomy, UCLA, Los Angeles, CA 90095, USA}
\affiliation{Mani L. Bhaumik Institute for Theoretical Physics, Department of Physics and Astronomy, UCLA, Los Angeles, CA 90095, USA}

\author{Bradley M. S. Hansen}
\affiliation{Department of Physics and Astronomy, UCLA, Los Angeles, CA 90095, USA}
\affiliation{Mani L. Bhaumik Institute for Theoretical Physics, Department of Physics and Astronomy, UCLA, Los Angeles, CA 90095, USA}

\begin{abstract}

Triple body systems are prevalent in nature, from planetary to stellar to supermassive black hole scales. In a hierarchical triple system, oscillations of the inner orbit's eccentricity and inclination can be induced on secular timescales. Over many cycles, the octupole-level terms in the secular equations of motion can drive the system to extremely high eccentricities via the Eccentric Kozai-Lidov (EKL) mechanism. The overall decrease in the inner orbit's pericenter distance has potentially dramatic effects for realistic systems, such as tidal disruption events. We present an analytical approximation in the test particle limit to describe individual step-wise increases in eccentricity of the inner orbit. A second approximation, also in the test particle limit, is obtained by integrating the equations of motion and calibrating to numerical simulations to estimate the overall octupole-level time evolution of the eccentricity. The latter approach is then extended beyond the test particle to the general case. The three novel analytical approximations are compared to numerical solutions to show that the models accurately describe the form and timescale of the secular descent from large distances to a close-encounter distance (e.g., the Roche limit). By circumventing the need for numerical simulations to obtain the long-term behavior, these approximations can be used to readily estimate properties of close encounters and descent timescales for populations of systems. We demonstrate by calculating rates of EKL-driven migration for Hot Jupiters in stellar binaries.
\end{abstract}

\section{Introduction} \label{sec:intro}

Hierarchical triple systems have been studied extensively in the literature in recent years, with a wide range of applications, from exoplanets to gravitational wave emitting black holes \citep[for a review, see][]{Naoz16}. A hierarchical system is defined by having a relatively tight ``inner'' binary with semi-major axis $a_1$ orbited by a faraway ``outer'' companion with semi-major axis $a_2$. In this case, the secular approximation can be applied, where the three-body Hamiltonian can be averaged over the orbital periods and expanded in powers of the small semi-major axis ratio $\alpha = a_1/a_2$ \citep[e.g.,][]{Kozai62, Harrington68, Ford+00, Naoz+13}. At the quadrupole level (proportional to $\alpha^2$), the inner orbit can undergo oscillations of eccentricity and inclination on timescales much longer than the orbital periods \citep{Kozai62, Lidov62}. The quadrupole level of approximation is often insufficient when the outer companion's orbit is eccentric, or when members have non-negligible mass \citep{Naoz+13}. In these cases, the octupole contribution (proportional to  $\alpha^3$) can cause the inner orbit to reach extremely high eccentricities or even flip from prograde to retrograde with respect to the total angular momentum, a phenomenon known as the Eccentric Kozai-Lidov (EKL) mechanism \citep{Naoz16}. 

These secular perturbations can lead the system to reach extreme eccentricities over a wide range of the parameter space \citep[e.g.,][]{Teyssandier+13,Li+14a,Li+14b}. Specifically, \citet{Li+14a} showed that the parameter space can be roughly divided into a chaotic regime and a nearly regular regime. In the latter, the long-term evolution of the quadrupolar cycles is modulated at the octupole level \citep[e.g.,][]{Naoz+11,Katz+11,Lithwick+11}. The octupolar envelope of the oscillations traces successive maxima of eccentricity reached during individual cycles. As the eccentricity increases, the pericenter distance of the inner orbiting body decreases. Therefore, the overall evolution of the inner orbiting body can be described as a \textit{secular descent} driven by the octupole-level terms.

Secular descents may play an important role in shaping planetary systems. In a system consisting of an inner binary with a star and Jupiter-like planet orbited by a distant planetary or stellar companion, secular perturbations from the distant body can excite the inner planet's orbital eccentricity, causing it to undergo close encounters with its host star \citep[e.g.,][]{Wu+03,Fabrycky+07,Naoz+11,Naoz+12,Li+14b,Petrovich+15a,Petrovich15b,Petrovich+16}. When tidal interactions are considered, the inner planet's semi-major axis can shrink and result in a Hot Jupiter \citep[e.g.,][]{Mayor+95, Naoz+11, Naoz+12, Petrovich+15a, Petrovich15b,Dawson+18}. Furthermore, the planet can be tidally disrupted if its pericenter distance is sufficiently reduced to cross the Roche limit, an effect that is thought to shape the demographics of the Hot Jupiter population \citep[e.g.,][]{Guillochon+11,Naoz+12, Valsecchi+15, Petrovich+15a, Munoz+16}.

The EKL mechanism may also drive secular descents in supermassive black hole (SMBH) binaries \citep[e.g.,][]{Li+15,Chen+09,Chen+11}. The outer SMBH can drive a star towards the inner SMBH to produce a tidal disruption event (TDE). EKL was shown to produce a TDE rate consistent with E+A (post-starburst) galaxies, as well as naturally produce partial, repeated TDEs \citep{Melchor+23}. Furthermore, this mechanism may be responsible for off-nuclear TDEs \citep{Mockler+23} and may also act to produce extreme-mass ratio inspiral gravitational wave events \citep{Naoz+22,Naoz+23}.

In each of these physical systems, it is necessary to understand the secular reduction of the inner orbit's pericenter from large values to a close-encounter distance (e.g., the Roche limit). While the eccentricity evolution can be numerically solved given the initial orbital configuration of a system, an analytical approximation for the long-term eccentricity evolution would be useful to readily estimate the form and timescale of the secular descent. For a large population of systems, having an analytical approximation would circumvent the need for costly numerical integrations and allow for efficient predictions of migration rates and demographic features.

In this paper, we derive an analytical model that describes the secular descent to some minimum pericenter distance. Specifically, we approach the problem from three different directions that provide results consistent with numerical simulations. In \S \ref{sec:picture}, we review the physical picture of the EKL mechanism to motivate the models. We present the models for both the test particle case and general case in \S \ref{sec:models}. Specifically, in \S \ref{sec:TP_swing} we employ a step-like approach to the changes in the z-axis angular momentum. Then, in \S \ref{sec:TP} and \ref{sec:beyondTP}, we use the equation of motions of the test particle and non-test particle, respectively, to derive the descent timescale. We describe the regimes of applicability for the models and quantify the accuracy of the models by comparing them to numerical simulations. Finally,  \S \ref{sec:applications} demonstrates the application of the models to the Hot Jupiter population. We summarize our conclusions in \S \ref{sec:conclusion}.

\section{Physical Picture and Motivation}
\label{sec:picture}

Here, we review the physical picture of 
the Eccentric Kozai-Lidov Mechanism \citep[see, e.g.,][]{Naoz16}. The hierarchical three-body system consists of an inner binary (masses $m_1$ and $m_2$) and a distant third body (mass $m_3$). We define the semi-major axis, eccentricity, argument of periastron, and the longitude of ascending nodes of the inner (outer) orbit as $a_1,e_1,\omega_1$ and $\Omega_1$ ($a_2,e_2,\omega_2$ and $\Omega_2$). We choose the z-axis to be parallel to the total angular momentum; thus, the longitude of ascending nodes have the following relations: $\Omega_1-\Omega_2=\pi$ \citep[e.g.,][]{Naoz+13}. 
The specific angular momentum is defined by $J_1=\sqrt{1-e_1^2}$ ($J_2=\sqrt{1-e_2^2}$) for the inner (outer) orbit.  In this case, the inclination of the inner and outer orbits is defined by the angular momentum components projected on the z-axis. In other words:
\begin{eqnarray}
   J_{z,1}  &=& J_1 \cos i_1 \\ 
   J_{z,2}  &=& J_2 \cos i_2 \ .
\end{eqnarray}
The mutual inclination is then simply $i=i_1+i_2$ \citep[e.g.,][]{Ford+00}. 

In order to motivate the analytical approximation below, we begin by describing the test particle system, which is a 2-degree-of-freedom system and, thus, provides a logical first step to analyze the physical system \citep[see also][]{Lithwick+11,Katz+11,Li+14a,Li+14b,Klein+23,Klein+24,Klein+24b,Klein+24c}. Under the test particle approximation ($m_2 \to 0$), the outer magnitude and organization of the angular momentum remain constant, where $i_2\to 0$ and $i_1\to i$, thus we define $\theta=\cos i_1$ and $J_z=J_{z,1}$. We also define $J=J_1$. In this case, the Hamiltonian can be written as \citep{Naoz16}: 
\begin{equation}
    \mathcal{H}^{T P}=\frac{3}{8} k^2 \frac{m_1 m_3}{a_2}\left(\frac{a_1}{a_2}\right)^2 \frac{1}{\left(1-e_2^2\right)^{3 / 2}}F \ ,
\end{equation}
where $k^2$ is Newton's constant, $F \equiv F_{\text {quad }}+\epsilon F_{\text {oct }}$ is the energy function, and
\begin{equation}
    \epsilon = \frac{a_1}{a_2} \frac{e_2}{1-e_2^2}
\end{equation}
characterizes the strength of the octupole component. The quadrupole component of $F$ is 
\begin{equation}
    F_{\text {quad }}=-\frac{e_1^2}{2}+\theta^2+\frac{3}{2} e_1^2 \theta^2+\frac{5}{2} e_1^2\left(1-\theta^2\right) \cos \left(2 \omega_1\right) \ ,
\end{equation}
and the octupole component is
\begin{equation}
    \begin{aligned} F_{\text {oct}} = & \frac{5}{16}\left(e_1+\frac{3}{4} e_1^3\right)\left[\left(1+11 \theta-5 \theta^2-15 \theta^3\right) \cos (\omega_1+\Omega_1)\right. \\ & \left.+\left(1-11 \theta-5 \theta^2+15 \theta^3\right) \cos (\omega_1-\Omega_1)\right] \\ & -\frac{175}{64} e_1^3\left[\left(1-\theta-\theta^2+\theta^3\right) \cos (3 \omega_1-\Omega_1)\right. \\ & \left.+\left(1+\theta-\theta^2-\theta^3\right) \cos (3 \omega_1+\Omega_1)\right] \ .\end{aligned}
    \label{eq:Foct}
\end{equation}

At the quadrupole level ($\epsilon=0$, corresponding to a circular outer orbit), $F$ and $J_z$ are the two constants of motion. There are two classes of trajectories, librating and circulating. Librating trajectories are associated with bound oscillations of $\omega_1$. The eccentricity evolves to large values on circulating trajectories, which have $e_1$ smallest and $i$ largest at $\omega_1 = 0$ and vice versa for $\omega_1 = \pm \pi/2$. The separatrix separates these modes of behavior and has $e_{1,0} = 0$ and $\theta = \sqrt{3/5}$. As such, large eccentricity oscillations are expected to occur between inclination angles of $\cos^{-1} \sqrt{3/5} \approx 39^{\circ}$ and $141^{\circ}$. 

For circulating trajectories with $e_{1,0} \ll 1$ and inclinations between $39^{\circ}$ and $141^{\circ}$, the maximum eccentricity $e_{1,\text{max}}$ is given by \citep[e.g.,][]{Lithwick+11}
\begin{equation} \label{eq:e1max}
    e_{1,\text{max}}^2 \approx 1 - \frac{5}{3}J_z^2 \ ,
\end{equation}
and the minimum eccentricity $e_{1,\text{min}}$ is given by
\begin{equation} \label{eq:e1min}
    e_{1,\text{min}}^2 \approx \frac{1}{2}(F-J_z^2) \ .
\end{equation}
\cite{Antognini15} estimated the timescale for each quadrupolar oscillation of $e_1$ to be\footnote{This is the general form. For the test particle case, $m_2 \to 0$.} 
\begin{equation}
    t_{\mathrm{quad}} \approx \frac{16}{15} \frac{a_2^3\left(1-e_2^2\right)^{3 / 2} \sqrt{m_1+m_2}}{a_1^{3 / 2} m_3 k} \ .
    \label{eq:t_quad}
\end{equation} 

\begin{figure}
\begin{center}

\includegraphics[width=3.5in]{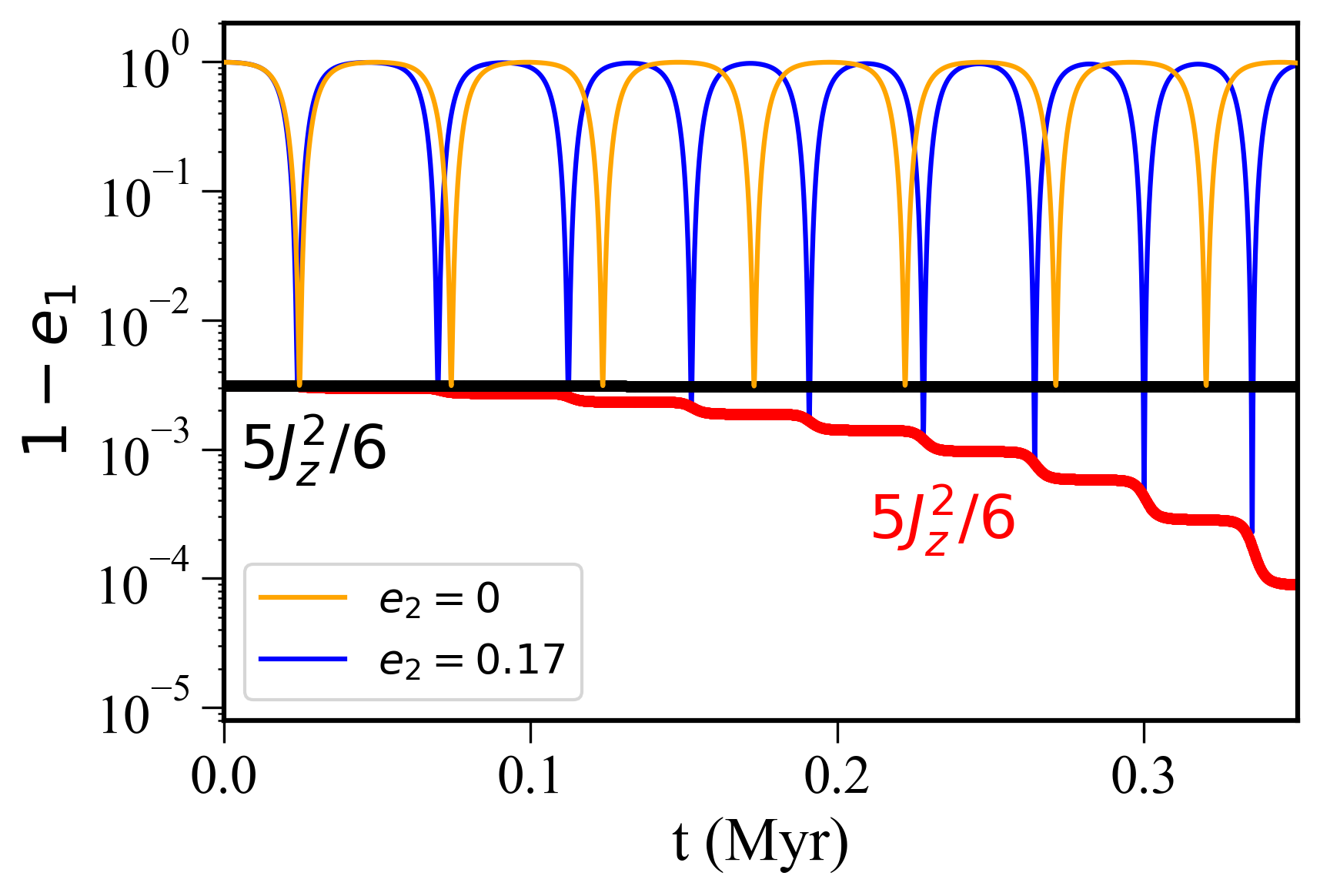}
\caption{\footnotesize Time evolution of $1-e_1$ for trajectories with $e_2 = 0$ (orange) and $e_2 = 0.17$ (blue) integrated numerically. Increasing the outer orbit's eccentricity drives $e_1$ to evolve to large values over multiple quadrupole cycles. $J_z$ controls the envelope of the eccentricity evolution, being constant in the $e_2 = 0$ case (black), and evolving in a step-wise fashion for $e_2>0$ (red).}
\label{fig:Jz_examples}
\end{center}
\end{figure}

For an eccentric outer body ($\epsilon>0$), $F$ is now the only conserved quantity, as the octupole component modulates $J_z$. Trajectories can be regular or chaotic \citep[e.g.,][]{Lithwick+11,Li+14b}; in this work, we focus on regular trajectories. For these, $J_z$ changes in a step-wise fashion, swinging to a new value during each plunge in $e_1$. The overall evolution of $J_z$ controls the envelope of the $e_1$ maxima. We demonstrate this behavior in Figure \ref{fig:Jz_examples}. It is this envelope that we seek to model, as it captures the overall {\it secular descent} of the inner body's pericenter distance $a_1 (1-e_1)$.

Relaxing the test particle approximation can lead to differences in orbital evolution. The inner orbit can torque the outer orbit, making large eccentricity oscillations occur in different regions of the parameter space \citep[e.g.,][]{Teyssandier+13,Naoz+14,Martin+15}. Below, we provide an analytical description for systems extending beyond the test particle limit. 

\section{Analytical Approximations of the Secular Descent}
\label{sec:models}

Here, we pedagogically describe our analytical approximations of the secular descent. They are divided into two cases: test particle and general. 

\subsection{Test particle approach}
\label{sec:testparticle}

As mentioned above, the test particle approach provides a 2-degree-of-freedom physical system. In the test particle approximation, $m_2\to 0$. The full set of equations can be found in \citet{Naoz16}. Here, in  Appendix \ref{app:EOM}, we present the equations of motion for completeness. We find that the analytical description can be further divided into two cases: one to describe single steps in $J_z$ and the other to describe the overall secular descent of the pericenter distance as eccentricity increases.

\subsubsection{Test particle approach, steps in eccentricity}
\label{sec:TP_swing}

We present an approach in the test particle limit to describe single steps in $J_z$. Here, we assume small initial $J_z$ ($J_{z,0} \lesssim 0.1$), corresponding to initially high mutual inclination ($i_0 \gtrsim 85^{\circ})$ for an initially circular orbit. The individual steps can be pieced together to estimate the overall secular descent, as illustrated in Figure \ref{fig:Jz_examples}. 

Since most of the change in $J_z$ occurs when $\dot{\omega}_1 \sim 0$ \citep[e.g.,][]{Li+14b}, in the limit of small $J_z$, $\dot{\omega}_1 \sim 0$ implies that (see Eq.~(\ref{eq:dom1dt_TP}))
\begin{equation}\label{eq:TPomegadot0}
    \cos{2 \omega_1} = \frac{1}{5} \frac{5J_z^2 - J^4}{J_z^2 - J^4} \approx \frac{1}{5} \ .
\end{equation}
We seek to characterize the change in $J_z$ during a single quadrupolar swing in $J$ (i.e. half a quadrupole cycle). We find the time evolution of $J$ from Eq.~(\ref{eq:dJdt_TP}), which for $\dot{\omega}_1 \sim 0$, from Eq.~(\ref{eq:TPomegadot0}), is 
\begin{equation} \label{eq:Jdot}
    \frac{dJ}{dt}\bigg|_{\dot{\omega}_1\sim 0} \approx -2 \sqrt{6} (1-J^2) \left[1 - \left(\frac{J_z}{J}\right)^2 \right] \ .
\end{equation}
Since $J_z$ at the quadrupole level is constant (for a test particle), we find the octupole level time evolution of $J_z$ from Eq.~(\ref{eq:dJzdt_TP}), which for $\dot{\omega}_1 \sim 0$ and taking the case $\Omega_1=0$ is
\begin{eqnarray}
    \frac{dJ_z}{dt}\bigg|_{\dot{\omega}_1\sim 0} &\approx& -\frac{5}{32} \sqrt{\frac{2}{5}} \epsilon \sqrt{1-J^2} \left(7-3 J^2\right)\frac{J_z}{J} \left(11-15 \frac{J_z^2}{J^2}\right) \nonumber \\& +&  \frac{245}{32}\sqrt{\frac{2}{5}}  \epsilon\left(1-J^2\right)^{3 / 2}\frac{J_z}{J} \left(1-\frac{J_z^2}{J^2}\right) \ .
\end{eqnarray}
To characterize the step in $J_z$ during a quadrupole cycle,
we divide $d{J_z}/dt$ by $dJ/dt$, which gives 
\begin{eqnarray}
    \frac{d J_z}{J_z} \approx \epsilon \frac{dJ}{J}  \left[\vphantom{\frac15} \right.  \frac{5}{64 \sqrt{15}} \frac{\left(7-3 J^2\right) }{\sqrt{1-J^2}}
    \left(11-15 \frac{J_z^2}{J^2}\right) \times \nonumber \\ \left(1-\frac{J_z^2}{J^2}\right)^{-1} - \frac{245}{64\sqrt{15}} \sqrt{1-J^2} \left. \vphantom{\frac12}\right] \ .
\end{eqnarray}

At face value, integrating both sides is challenging because $J_z$ appears both in the left and right-hand side of the equation. However, recall that over a quadrupole cycle, $J$ changes significantly, but $J_z$ remains approximately constant \citep{Lithwick+11}. Thus, we treat $J_z$ as a constant and find the approximation
\begin{eqnarray}
    \ln J_z &\approx & \frac{35}{16 \sqrt{15}} \epsilon \ln \left( {\frac{1-\sqrt{1-J^2}}{{1+\sqrt{1-J^2}}}} \right) \nonumber \\   & + & \frac{5}{32 \sqrt{15}} \epsilon \frac{(7-3J_z^2)}{\sqrt{1-J_z^2}} \ln \left(  \frac{-\sqrt{1-J^2}+\sqrt{1-J_z^2}}{\sqrt{1-J^2}+\sqrt{1-J_z^2}} \right) \nonumber \\  &-& \frac{5}{4\sqrt{15}} \epsilon \sqrt{1-J^2}+\text {const. } \ .
\end{eqnarray}
From here we find the relation between the final ($J_{z,f}$) and initial ($J_{z,0}$) values of $J_z$ during a step in $J$ (half a quadrupole cycle) to be    
\begin{eqnarray} 
    \frac{J_{z,f}}{J_{z,0}}&\approx&\exp \left(- \frac{5}{4\sqrt{15}} \epsilon \sqrt{1-J^2}\right)  \nonumber \\ &\times& \left( {\frac{1-\sqrt{1-J^2}}{{1+\sqrt{1-J^2}}}} \right)^{\zeta_1}  \nonumber \\ &\times& \left(  \frac{-\sqrt{1-J^2}+\sqrt{1-J_z^2}}{\sqrt{1-J^2}+\sqrt{1-J_z^2}} \right)^{\zeta_2} \ ,
\label{eq:Jz_swing}
\end{eqnarray}
where
\begin{equation}
    \zeta_1 = \frac{35}{16\sqrt{15}} \epsilon \ ,
\end{equation}
and
\begin{equation}
    \zeta_2 = \frac{5}{32\sqrt{15}} \epsilon \frac{(7-3J_z^2)}{\sqrt{1-J_z^2}} \ .
\end{equation}
We use this expression to approximate the step in $J_z$ from low to high eccentricity. For the step from high to low eccentricity, the system swings to the other $\dot{\omega_1}= 0$ root, reversing the sign of $dJ/dt$.

\begin{figure*}
\begin{center}

\includegraphics[width=7in]{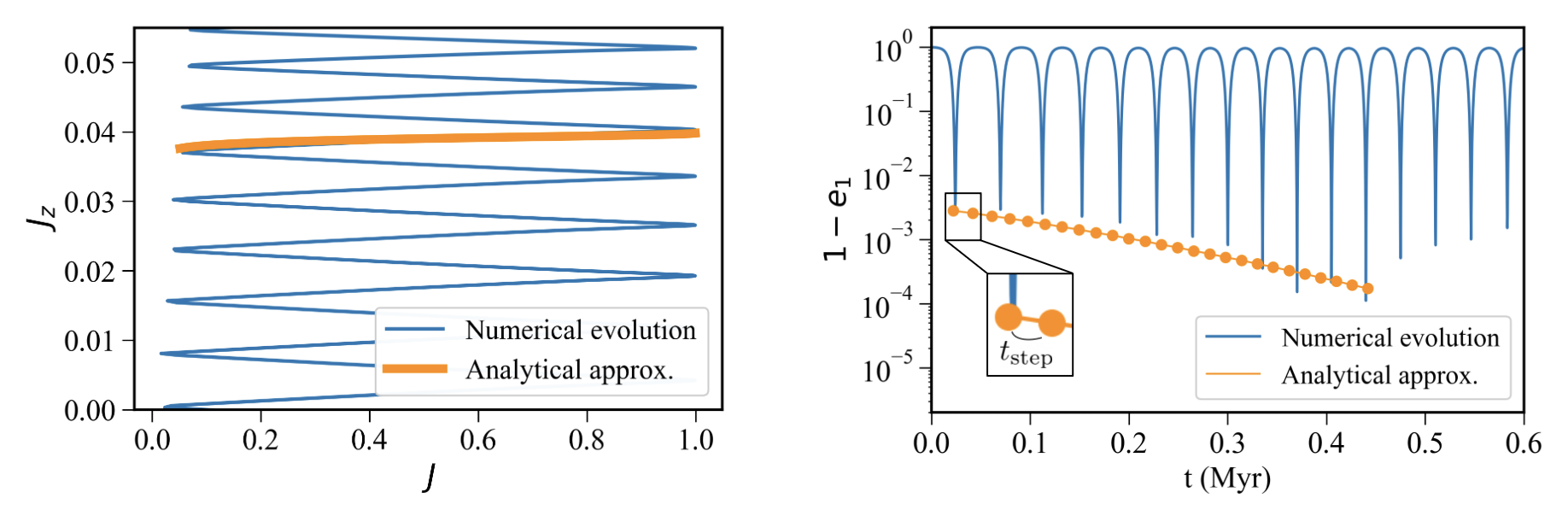}
\caption{\footnotesize Comparison of the numerical evolution to the analytical approximation for the step-wise eccentricity evolution. Here we consider a test particle at 1 AU from a $1M_{\odot}$ star orbited by a distant perturber of mass $0.16M_{\odot}$ at 20 AU. The system is initially set with $e_1 = 0.01$, $e_2=0.17$, $\omega_1 = 0^{\circ}$, $\omega_2 = 0^{\circ}$, and $i=86.5^{\circ}$. Left: We compare one step in $J_z$ (orange) modeled by Eq.~(\ref{eq:Jz_swing}) to the numerical evolution (blue). We find that the evolution of $J_z$ closely agrees for much of the step. The lower accuracy at low $J$ is explained by $\omega_1$ varying here and the assumption of $\dot{\omega}_1 = 0$ breaking down.  Right: We stitch together the individual steps (orange dots represent each step, the orange curve represents the overall secular descent) as described in \S \ref{sec:TP_swing}. The zoom-in panel shows one step with timescale estimated from Eq.~(\ref{eq:t_step}). Comparing to the numerical evolution (blue), we see that the overall time evolution of $J_z$ is closely captured.}
\label{fig:Jz_model_example}
\end{center}
\end{figure*}

The timescale $t_{\text{step}}$ between a local maximum and a minimum in $J$ can then be found by integrating Eq.~(\ref{eq:Jdot}) between the maximum eccentricity in Eq.~(\ref{eq:e1max}) and the minimum eccentricity in Eq.~(\ref{eq:e1min}), giving
\begin{multline} \label{eq:t_step}
 t_{\text{step}} \approx  \\ 
 \int_{\sqrt{1-\frac{1}{2}(F-J_z^2)}}^{\sqrt{\frac{5}{3}}J_z} \left( -2 \sqrt{6} [1-J^2] \left[1 - \left(\frac{J_z}{J}\right)^2 \right]\right)^{-1} dJ  \\ =
 -\frac{1}{4\sqrt{6}(J_z^2-1)} \left[\vphantom{\frac12} \right. J_z\ln( |J+J_z | ) -J_z \ln( |J-J_z | ) \\ - \ln(|J+1|) + \ln(|J-1|) \left. \vphantom{\frac12}\right] \bigg|_{\sqrt{1-\frac{1}{2}(F-J_z^2)}}^{\sqrt{\frac{5}{3}}J_z}
 \ , 
\end{multline}
where the conversion to true time is described in Eq.~(\ref{eq:truetime}). See the annotation on the right-hand side of Figure \ref{fig:Jz_model_example}. We note that this timescale is similar to half the quadrupole timescale given in Eq.~(\ref{eq:t_quad}), but it is more accurate for our needs. 

The total timescale for a descent then is 
\begin{equation}
    t_{\rm descent, TP} = \sum_{J_{z,0}}^{J_{z, \rm min}} t_{\rm step} \ ,
\end{equation}
where $J_z$ is updated on each step from Eq.~(\ref{eq:Jz_swing}). $J_{z,\rm min}$ can be set at a desired minimum pericenter distance (e.g., the Roche limit). In a realistic system, the octupole-level of approximation can drive the eccentricity to extreme values, which is often associated with an orbital flip \citep[e.g.,][]{Naoz+11,Katz+11,Lithwick+11,Li+14a,Teyssandier+13}. We find that for systems that flip, the octupole level maximum eccentricity reached at the flip is roughly at the range of $1-e_1 \sim 10^{-4}$ to $10^{-6}$, which agrees with values obtained in previous investigations \citep[e.g.,][]{Katz+11,Li+14b,Liu+15}. In this work, we specify the maximum eccentricity that we allow the system to evolve to. 

In the left panel of Figure \ref{fig:Jz_model_example}, we show the changes of $J_z$ as a function of $J$ for a representative system calculated numerically \citep[][]{Naoz+13}, overplotting the analytical model in Eq.~(\ref{eq:Jz_swing}). In the right panel, we present the time evolution of $1-e_1$ of the same system, overplotting the analytical secular descent as a function of time. To plot this line, we stitch together individual $J_z$ steps and estimate the timescale for each step from  Eq.~(\ref{eq:t_step}). The analytical model and numerical evolution agree to within a small factor. Therefore, this analytical approximation can be used to closely estimate the timescale to reach some eccentricity maximum, as well as the step-wise increase of the pericenter distance.

One application of this model is in studies of tidal disruption, in which it is necessary to understand the ratio of the Roche limit to the pericenter distance. For systems in which this ratio is higher, stronger tidal interactions are more likely to occur \citep[e.g.,][]{Guillochon+11}. At high eccentricity ($J \to 0$), the pericenter distance is small and the system nears the Roche limit. In this regime, the analytical model of Eq.~(\ref{eq:Jz_swing}) for successive jumps in $J_z$ and the relation between $J_z$ and eccentricity in Eq.~(\ref{eq:e1max}) can be combined to understand the pericenter evolution. For a system with $\epsilon = 0.01$, changes in the pericenter distance with each successive quadrupole cycle are $\sim$10\%, so that the first tidal encounters will be grazing. For a system with higher $\epsilon$, the first tidal encounter may be well within the Roche limit. As such, the analytical model in this work can provide an estimate for how much the pericenter distance increases between successive quadrupole cycles due to secular perturbations, providing insight into the nature of close encounters.

The strength of this analytical model lies in this ability to estimate the changes in eccentricity during each cycle, which can be pieced together to approximate the full secular descent. However, this approach is restricted to the small $J_z$ limit. For systems on initially circular orbits, this corresponds to near-perpendicular initial mutual inclinations. Next, we take a different approach that extends beyond this regime and is able to describe the entire descent of the pericenter distance without the need to piece individual steps together.

\subsubsection{Test particle approach, full descent}
\label{sec:TP}

Here, we analytically describe the time evolution of a test particle's maximum eccentricity over a full secular descent. This approach differs from the approach in \S \ref{sec:TP_swing}, in which we focused on regions where $e_1$ most rapidly changes ($\omega_1 \sim 39^{\circ}$) to estimate the change in $J_z$ during an individual quadrupole cycle. Here, we focus on connecting the points where the eccentricity is largest ($\omega_1 \sim 90^{\circ}$) to trace the pericenter evolution. 
 Therefore, we work with the maximum eccentricity, $e_{1,\text{max}}$, to analytically describe the envelope of maximum eccentricity. We adopt the test particle equation of motion and integrate to obtain a single analytical formula for the overall secular descent for systems with $i_0 \gtrsim 45^{\circ}$. This approach is motivated by \cite{Katz+11}, in which the equations of motion are averaged over the quadrupole cycles to yield approximate equations for the long-term octupolar eccentricity evolution. 

\begin{figure*}
\begin{center}

\includegraphics[width=5in]{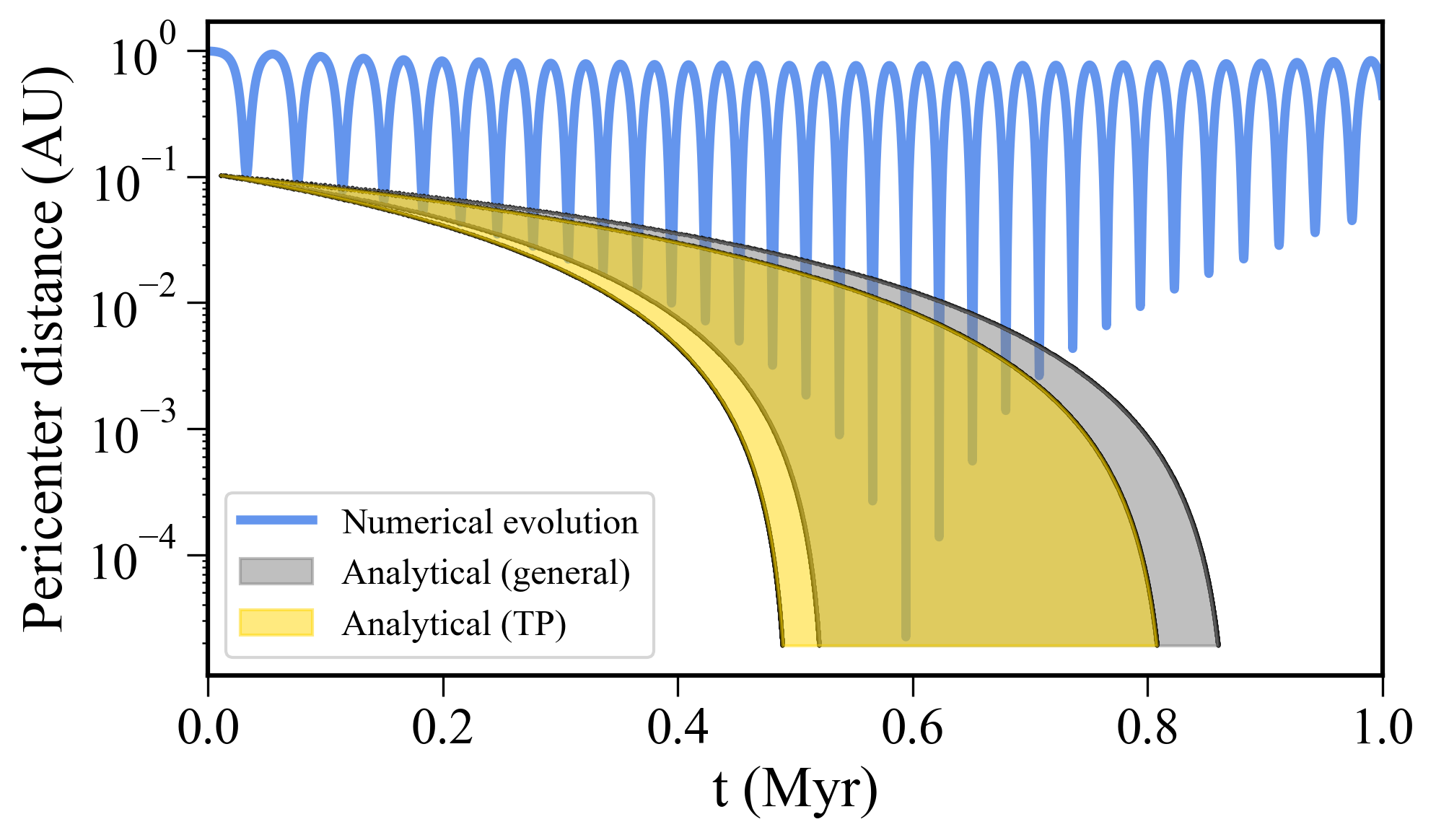}
\caption{\footnotesize Comparison of the numerical evolution (blue) with the analytical approximations to the secular descent for an example Jupiter-like planet. The gold-shaded region corresponds to the test particle case from Eq.~(\ref{eq:t_TP}), and the grey-shaded region corresponds to the general case from Eq.~(\ref{eq:t_nonTP}) (see \S \ref{sec:beyondTP}). The spread in the shaded regions represents the $\sim$25\% variations in the numerical factor $\Upsilon$ obtained from Monte Carlo simulations (see Appendix \ref{app:fitting}), corresponding to varying evolutions of $\omega_2$. Here we consider a $1M_{\text{J}}$ planet at 1 AU from a $1M_{\odot}$ star orbited by a distant perturber of mass $0.1M_{\odot}$ at 20 AU. The system is initially set with $e_1 = 0.01$, $e_2=0.4$, $\omega_1 = 0^{\circ}$, $\omega_2 = 0^{\circ}$, and $i=70^{\circ}$. The approximations closely track the envelope of the eccentricity oscillations through the descent.}
\label{fig:examplesystem}
\end{center}
\end{figure*}

Here, we assume small initial eccentricity\footnote{This approximation holds, roughly as long as $e_{1,0}\lesssim 0.3$, \citep[e.g.,][]{Li+14b}. }, $e_{1,0} \ll 1$. In this case, the maximum eccentricity for a single quadrupole cycle is approximated by Eq.~({\ref{eq:e1max}}). Differentiating the expression with respect to $t$, we obtain
\begin{equation}
      \frac{d{e_{1,\text{max}}}}{dt} \approx -  \frac{5}{3} \theta \frac{1}{e_{1,\text{max}}} \sqrt{1-e_{1,\text{max}}^2} \frac{dJ_z}{dt} \ .
\end{equation}
The quadrupole component of ${dJ_z}/{dt}$ is zero, and the octupole component is obtained from Equations (\ref{eq:Foct}) and (\ref{eq:dJzdt_TP}). For trajectories in which $\omega_1$ is circulating, the eccentricity is largest at $\omega_1 \sim \pm \pi/2$ and $\theta \sim \sqrt{3/5}$. Thus, plugging in these values, we obtain an expression for the time evolution of $e_{1,\text{max}}$:
\begin{eqnarray} \label{eq:dedt}
      \frac{d{e_{1,\text{max}}}}{dt} f({e_{1,\text{max}}}) &\approx& \nonumber \\
       \frac{15}{64} k \frac{a_1^{5/2}}{a_2^4} \frac{m_3}{\sqrt{m_1}} \frac{e_2}{(1-e_2^2)^{5/2}} \cos{\omega_2} && \ ,
\end{eqnarray}
where 
\begin{eqnarray}
    f({e_{1,\text{max}}}) = \frac{1}{(2+5e_{1,\text{max}}^2)\sqrt{1-e_{1,\text{max}}^2}} \ .
\end{eqnarray}
We are interested in the pericenter distance of the inner body and thus adopt $r_p = 1 - e_{1,\text{max}}$. Therefore, we can write Eq.~(\ref{eq:dedt}) as
\begin{equation} \label{eq:drpdt_TP}
\frac{d{r_p}}{dt} f(r_p) \approx -\frac{15}{64} k \frac{a_1^{5/2}}{a_2^4} \frac{m_3}{\sqrt{m_1}} \frac{e_2}{(1-e_2^2)^{5/2}} \cos{\omega_2}  \ ,
\end{equation}
where $f(r_p)$ is expanded by taking $r_p \ll 1$ 
\begin{eqnarray}
f(r_p)=\frac{1}{(5r_p^2-10r_p+7)\sqrt{-r_p^2+2r_p} } \approx \nonumber \\ 
\frac{1}{7\sqrt{2}}\frac{1}{\sqrt{r_p}} + \frac{47}{196\sqrt{2}}\sqrt{r_p} +  \frac{2787}{10976\sqrt{2}}r_p^{3/2} + \nonumber \\ \left(\frac{3165}{43904\sqrt{2}}+ \frac{150\sqrt{2}}{2401}\right) r_p^{5/2} + \mathcal{O}(r_p^{7/2}) \ .  
\end{eqnarray}
Eq.~(\ref{eq:drpdt_TP}) can be integrated to obtain the time for a minimum $r_{p,\rm min}$. For the limits on the integral, the eccentricity at the first quadrupole maximum is estimated from Eq.~(\ref{eq:e1max}) and the quadrupole timescale to reach the first maximum from Eq.~(\ref{eq:t_quad}). One can stop the integration once a desired pericenter distance is reached, such as the Roche limit (see discussion of octupole level maximum eccentricity in \S \ref{sec:TP_swing}). The integral is
\begin{eqnarray} \label{eq:integral_TP}
    \Upsilon_{\rm TP} \int_{1-\sqrt{1 - \frac{5}{3}J_{z,0}^2}}^{r_{p, \rm min}} {d{r_p'}} f(r_p') \approx \nonumber \\ -\frac{15}{64} k \frac{a_1^{5/2}}{a_2^4} \frac{m_3}{\sqrt{m_1}} \frac{e_2}{(1-e_2^2)^{5/2}}  \int_{t_{\text{quad}}}^{t_{r_{p,\rm min}}} dt' \ ,
\end{eqnarray}
where $\Upsilon_{\rm TP}$ is a numerical factor that captures the evolution of $\omega_2$, see below. 
The solution of this integral provides the relevant timescale to descend to a minimum $r_p$ over an octupole cycle: 
\begin{equation}\label{eq:t_TP}
     t_{\rm descent} = t_{\text{quad}}  + \Upsilon_{\rm TP} t_{\rm oct} \eta(r_{p,\rm min}) \ , 
\end{equation}
where
\begin{equation} \label{eq:t_oct_TP}
    t_{\rm oct}=\frac{64}{15} k^{-1}\frac{a_2^4}{a_1^{5/2}} \frac{\sqrt{m_1}}{m_3} \frac{(1-e_2^2)^{5/2}}{e_2} \ ,
\end{equation}
and 
\begin{eqnarray} 
   \eta(r_{p,\rm min}) = \left[\vphantom{\frac12} \right. \frac{\sqrt{2}}{7}\sqrt{r_p}  + \frac{47}{294 \sqrt{2}} r_p^{3/2} + \frac{2787}{27440 \sqrt{2}} r_p^{5/2} +  \nonumber \\ \frac{60555}{1075648 \sqrt{2}} r_p^{7/2} + \mathcal{O}(r_p^{9/2})  \left. \vphantom{\frac12}\right] \bigg|_{r_{p, \rm min}}^{1-\sqrt{1 - \frac{5}{3}J_{z,0}^2}} \ .
\end{eqnarray} 
We note that the high-order terms do not contribute significantly to the accuracy of the approximation. Figure \ref{fig:examplesystem} shows a comparison of the numerical evolution with the analytical model for an example system.

To evaluate the numerical factor $\Upsilon_{\rm TP}$, we run 500 Monte Carlo realizations of systems, changing the mass ratio, semi-major axis ratio, inclination, and $e_2$ 
 (see Appendix \ref{app:fitting} for procedure). We find that the initial mutual inclination is the main parameter that affects the value of $\Upsilon_{\rm TP}$ (see the left panel in Figure \ref{fig:fitting}, in Appendix \ref{app:fitting}). Thus, given the initial mutual inclination, the descent to a minimum $r_p$ is well characterized. The variations in our estimation of $\Upsilon_{\rm TP}$ are $\sim$25\% for initial inclinations above $\sim$60$^{\circ}$, depicted as the shaded area in Figure \ref{fig:examplesystem}. 
 
\begin{figure}[]
\begin{center}

\includegraphics[width=\linewidth]{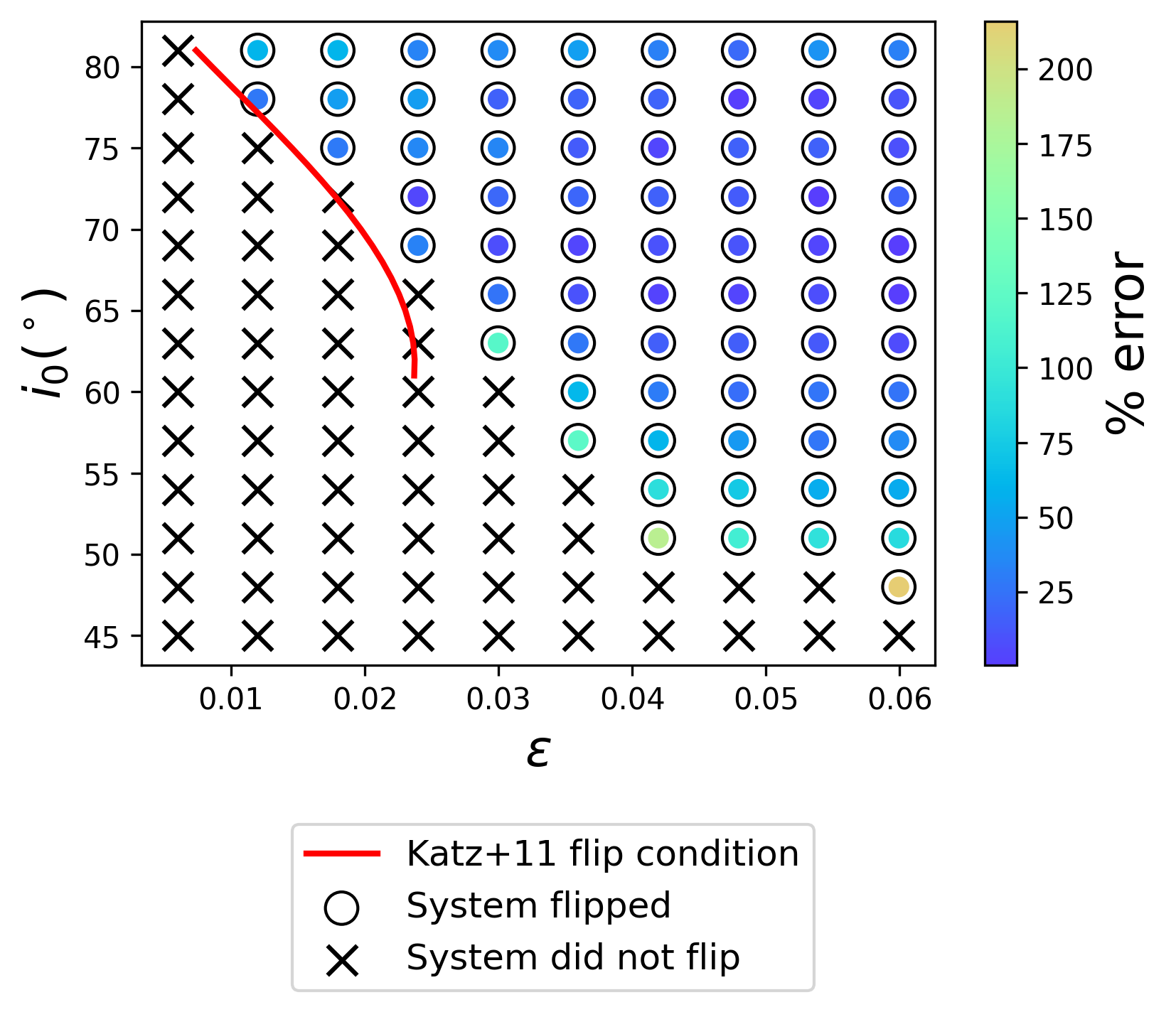}
\caption{\footnotesize Comparison of the numerical descent timescale with the analytical test particle approximation of Eq.~(\ref{eq:t_TP}) for varied $\epsilon$ and $i_0$. We start each system with $\omega_1=\omega_2=0$. Darker circles correspond to a lower percentage error in the timescale estimation. Circles show systems that flipped within 1000$t_{\rm quad}$ (thus displaying extreme eccentricity evolution), and we marked with "X" systems that did not flip. We also plot the flip condition from \cite{Katz+11} (red curve), which applies for systems with $i_0 \gtrsim 60^{\circ}$. We see that the flip condition accurately separates the systems that flip from those that did not. For the systems that flipped and underwent secular descents, the model matches the numerical result to within $25\%$ for most systems, as expected given the variations observed in the Monte Carlo simulations in Appendix \ref{app:fitting}. The model matches the numerical result to a factor of a few for all systems here that undergo a flip.}
\label{fig:ei_grid}
\end{center}
\end{figure}

 We show the percentage error between the numerical evolution and the model in Figure \ref{fig:ei_grid} for systematically varied initial $i_0$ and $\epsilon$ between $\sim$$45-80^\circ$ and $0.006-0.06$, respectively. Here, we fixed the masses and initial eccentricity of both orbits, and set $\omega_1 = \omega_2 = 0$ initially.  We estimated the percentage error between the model timescale depicted in Eq.~(\ref{eq:t_TP}) and the numerical descent, shown as the color code. We also over-plot the analytical flip condition from \citet{Katz+11}. We note that the flip of a system is closely associated with the octupolar excitation to extreme eccentricity, and the parameter space in which deep descents occur is very similar to the parameter space in which flips occur \citep[e.g.,][]{Lithwick+11,Katz+11}. In Figure \ref{fig:ei_grid}, we marked with ``X'' systems that did not flip within 1000 quadrupole timescales, and we marked with circles systems that did flip, thus displaying extreme eccentricity evolution. Notably, the novel model is able to accurately describe the evolution of systems that undergo extreme eccentricity evolution to within a small factor.

Deep descents in $r_p$ tend to occur when $\omega_1 = \omega_2  = 0$ or $\omega_1 = \omega_2  = \pi$, which is not surprising because these values are associated with the resonances \citep[e.g.,][]{Hansen+20}. Therefore, in the pedagogical examples, 
we set our systems nominally with the initial condition $\omega_1 = \omega_2  = 0$ to produce immediate descents. We later relax this condition and perform numerical studies that show that for most systems with sufficiently high $i$ and $\epsilon$, $\omega_1$ and $\omega_2$ will quickly synchronize to produce a descent if they are initially unsynchronized. We define the beginning of the descent in the numerical results by working backward from the maximum eccentricity until the first local eccentricity maximum having $e_1$ within $5 \times 10^{-3}$ of the quadrupole maximum eccentricity in Eq.~(\ref{eq:e1max}). We find that this criterion effectively captures the numerical deep descents from the local quadrupole maximum to a very large eccentricity for nearly every system.

\begin{figure*}
\begin{center}

\includegraphics[width=7.0in]{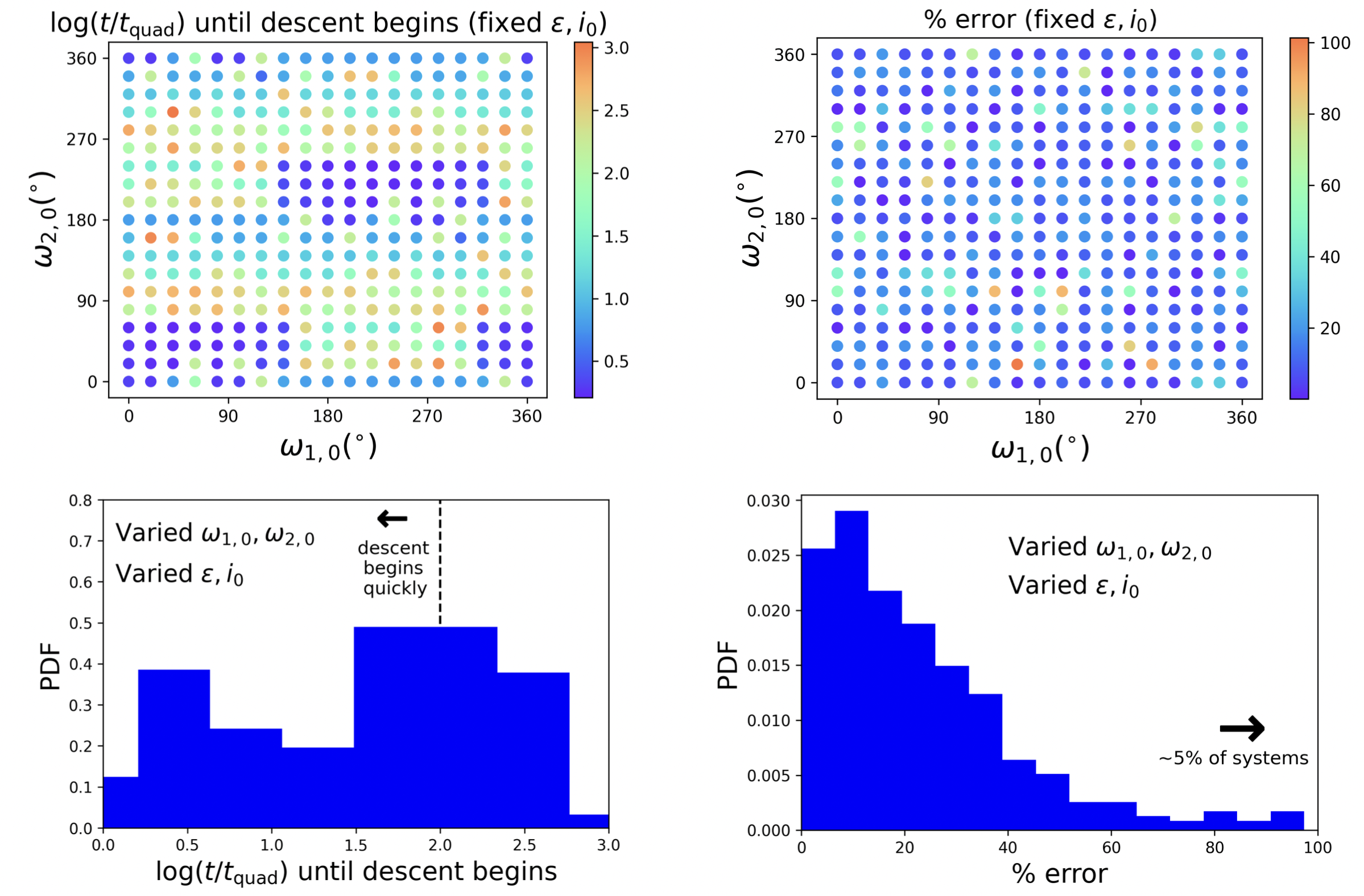}
\caption{\footnotesize Top left: Time until a deep descent in $1-e_1$ begins for varied  $\omega_{1,0}$ and $\omega_{2,0}$. We show the logarithm of the time in quadrupole cycles. Here, we define a deep descent as one reaching $1-e_1 = 10^{-3}$. Every system here has initially $m_1 = 1M_{\odot}, m_2 = 1M_J, m_3=0.1M_{\odot}, a_1 = 1$ AU, $a_2 = 20$ AU, $e_1 = 0.01$, $e_2 = 0.4$, and $i = 75^{\circ}$. Darker circles correspond to a more rapid time for the descent to begin.  We find that given sufficient time ($\sim$1000 quadrupole cycles), every system here eventually undergoes such a descent, with $\sim$60\% happening quickly (in $<$100 quadrupole cycles). Top right: For the same set of systems as in the top left, we compare the numerical timescale of the descent once it occurs to the analytical test particle approximation in Eq.~(\ref{eq:t_TP}). Darker circles correspond to lower percentage error in the descent timescale. We find that the model agrees with the numerical evolution to within $20\%$ for $>$80\% of systems and agrees within a factor of $\sim$2 for all systems. Bottom left: We vary $\omega_{1,0}$ and $\omega_{2,0}$ as in the top panels, while also varying $\epsilon$ from $0.02-0.09$ and $i_0$ from $60-80^{\circ}$. We plot the distribution of times until the descent begins. We again find every system eventually undergoes a descent, with $\sim$60\% happening in $<$100 quadrupole cycles. Bottom right: For the same set of systems as the bottom left panel, we plot the distribution of percentage errors comparing the numerical descent timescale to the analytical test particle approximation. We find that the model agrees with the numerical evolution to within $20\%$ for $\sim$50\% of systems and to within a factor of a few for all systems.}
\label{fig:omegagrid}
\end{center}
\end{figure*}

To highlight this behavior, we run two separate numerical tests. In one, we fix $i_0=75^\circ$ and $\epsilon=0.024$ and vary the initial $\omega_{1,0}$ and $\omega_{2,0}$ between $0-360^{\circ}$ along a grid. We first examine how long it takes for the descent to begin (as a function of the quadrupole timescale). This is shown in the panels of Figure \ref{fig:omegagrid}, wherein the top left color codes the timescale. We demonstrate that for a wide range of the parameter space, the descent begins within 100 quadrupole cycles. We also run a Monte Carlo as a proof of concept for systems with $i_0 = 60-80^{\circ}$ and $\epsilon = 0.02-0.09$, while varying the initial $\omega_{1,0}$ and $\omega_{2,0}$ between $0-360^{\circ}$ along a grid. These ranges were chosen to encompass systems within the flip parameter space of \cite{Katz+11}. The timescales for the Monte Carlo run are shown in the bottom left panel of Figure \ref{fig:omegagrid}. We find that the majority of systems start to descend within 100 quadrupole cycles. 

Given a long period of time, chaos may lead nearly all such systems to eventually produce a deep descent. Once the descent begins, we find that our model is generally able to closely capture the evolution to within a small factor. This is shown in the right-hand panels of Figure \ref{fig:omegagrid} where we use the same two numerical runs described above and show the percentage error between the numerical descent and the analytical model, Eq.~(\ref{eq:t_TP}). As depicted, for the majority of systems, the error is less than $20\%$, for a system with fixed $\epsilon$ and fixed $i_0$. We further estimate the difference between the numerical and analytical models for the Monte Carlo run with varied $\epsilon$ and $i_0$. Only $\sim$$5\%$ of the systems have an error larger than a factor of 2, and the majority of systems have an error smaller than 20\%.

\subsection{Beyond the test particle}
\label{sec:beyondTP}

Now we move beyond the test particle limit and consider the general case with nonzero $m_2$. We begin by taking the full expression for ${de_1}/{dt}$ given in Eq.~(\ref{eq:dedt_full}). Following the procedure in \S \ref{sec:TP}, we set $\omega_1 = \pi/2$ and $\theta = \sqrt{3/5}$ to capture the eccentricity maxima. We obtain
\begin{eqnarray} \label{eq:dedt_nonTP}
      \frac{d{e_{1,\text{max}}}}{dt} f({e_{1,\text{max}}}) &\approx& \nonumber \\
      \frac{15}{64} k \frac{a_1^{5/2}}{a_2^4} \frac{(m_1-m_2)m_3}{(m_1+m_2)^{3/2}} \frac{e_2}{(1-e_2^2)^{5/2}} \cos{\omega_2} && \ ,
\end{eqnarray}
where 
\begin{eqnarray}
    f({e_{1,\text{max}}}) = \frac{1}{(-2+9e_{1,\text{max}}^2) \sqrt{1-e_{1,\text{max}}^2}} \ .
\end{eqnarray}
As in \S \ref{sec:TP}, we adopt $r_p = 1 - e_{1,\text{max}}$ to write
\begin{equation} \label{eq:drdpdt_nonTP}
\frac{d{r_p}}{dt} f(r_p) \approx -\frac{15}{64} k \frac{a_1^{5/2}}{a_2^4} \frac{(m_1-m_2)m_3}{(m_1+m_2)^{3/2}} \frac{e_2}{(1-e_2^2)^{5/2}} \cos{\omega_2}  \ ,
\end{equation}
where $f(r_p)$ is expanded by taking $r_p \ll 1$
\begin{eqnarray}
f(r_p)=\frac{1}{(9r_p^2-18r_p+7)\sqrt{-r_p^2+2r_p} } \approx \nonumber \\ 
\frac{1}{7\sqrt{2}}\frac{1}{\sqrt{r_p}} + \frac{79}{196\sqrt{2}}\sqrt{r_p} + \frac{9507}{10976\sqrt{2}}r_p^{3/2} + \nonumber \\ \left(\frac{10109}{43904\sqrt{2}}+ \frac{1782\sqrt{2}}{2401}\right) r_p^{5/2} + \mathcal{O}(r_p^{7/2}) \ .
\end{eqnarray}
Eq.~(\ref{eq:drdpdt_nonTP}) can be integrated 
\begin{eqnarray} \label{eq:integral_beyondTP}
    \Upsilon \int_{1-\sqrt{1 - \frac{5}{3}J_{z,0}^2}}^{r_{p, \rm min}} {d{r_p'}} f(r_p') \approx \nonumber \\ -\frac{15}{64} k \frac{a_1^{5/2}}{a_2^4} \frac{(m_1-m_2)m_3}{(m_1+m_2)^{3/2}} \frac{e_2}{(1-e_2^2)^{5/2}} \int_{t_{\text{quad}}}^{t_{r_{p,\rm min}}} dt' \ ,
\end{eqnarray}
where $\Upsilon$ is a numerical factor that captures the evolution of $\omega_2$ as discussed in \S \ref{sec:TP}. The solution of the integral is the relevant timescale to descend to a minimum $r_p$ over an octupole cycle:
\begin{equation}\label{eq:t_nonTP}
     t_{\rm descent} = t_{\text{quad}}  + \Upsilon t_{\rm oct} \eta(r_{p,\rm min}) \ , 
\end{equation}
where
\begin{equation}
    t_{\rm oct}=\frac{64}{15} k^{-1}\frac{a_2^4}{a_1^{5/2}} \frac{(m_1+m_2)^{3/2}}{(m_1-m_2)m_3}  \frac{(1-e_2^2)^{5/2}}{e_2} \ , 
\end{equation}
and 
\begin{eqnarray} 
    \eta(r_{p,\rm min}) &=& \left[\vphantom{\frac12} \right. \frac{\sqrt{2}}{7}\sqrt{r_p}  + \frac{79}{294 \sqrt{2}} r_p^{3/2} + \frac{9507}{27440 \sqrt{2}} r_p^{5/2} +  \nonumber \\ &+& \frac{526955}{1075648 \sqrt{2}} r_p^{7/2} + \mathcal{O}(r_p^{9/2})  \left. \vphantom{\frac12}\right] \bigg|_{r_{p, \rm min}}^{1-\sqrt{1 - \frac{5}{3}J_{z,0}^2}} \ .
\end{eqnarray} 
In the test particle case (\S \ref{sec:TP}), the dependence of $t_{\rm oct}$ is similar; see Eq.~(\ref{eq:t_oct_TP}). We again note that the high-order terms do not contribute significantly to the accuracy of the approximation.

As in the test particle case, we fit to Monte Carlo simulations to estimate values of $\Upsilon$ (see Appendix \ref{app:fitting} for procedure). We again find that the initial mutual inclination is the main parameter that affects the value of $\Upsilon$ (see the right panel in Figure \ref{fig:fitting}, in Appendix \ref{app:fitting}). We find that the average variations in $\Upsilon$ are $\sim$25\%, similar to the test particle case. Figure \ref{fig:examplesystem} shows a comparison of the numerical evolution with the analytical model for an example system, also compared with the test particle approximation. We find that the two approaches provide consistent results.

\section{Application to Hot Jupiters}
\label{sec:applications}

Here, we demonstrate the application of our model to a population of astrophysical systems. We estimate the EKL-driven high-eccentricity migration timescale for Hot Jupiters in stellar binaries. We use the test particle approximation in Eq.~(\ref{eq:t_TP}) to estimate the time it takes for a Jupiter-like planet in a stellar binary to descend from large distances to the Roche limit, which we define as \citep[e.g.,][]{Guillochon+11}
\begin{equation}
    R_{\rm Roche} = 2.7 R_2 \left(\frac{m_1+m_2}{m_2}\right)^{1/3} \ ,
\end{equation}
where $R_2$ is the radius of the secondary body (in this case, the planet). The Roche limit is $\sim$0.01 AU for a Jupiter-like planet and Sun-like star, requiring an orbital eccentricity of $\sim$0.99 for a planet initially at 1 AU. In Appendix \ref{app:timescales}, we compare the descent timescale calculated with the model with other relevant physical timescales.

\begin{figure*}
\begin{center}

\includegraphics[width=3.5in]{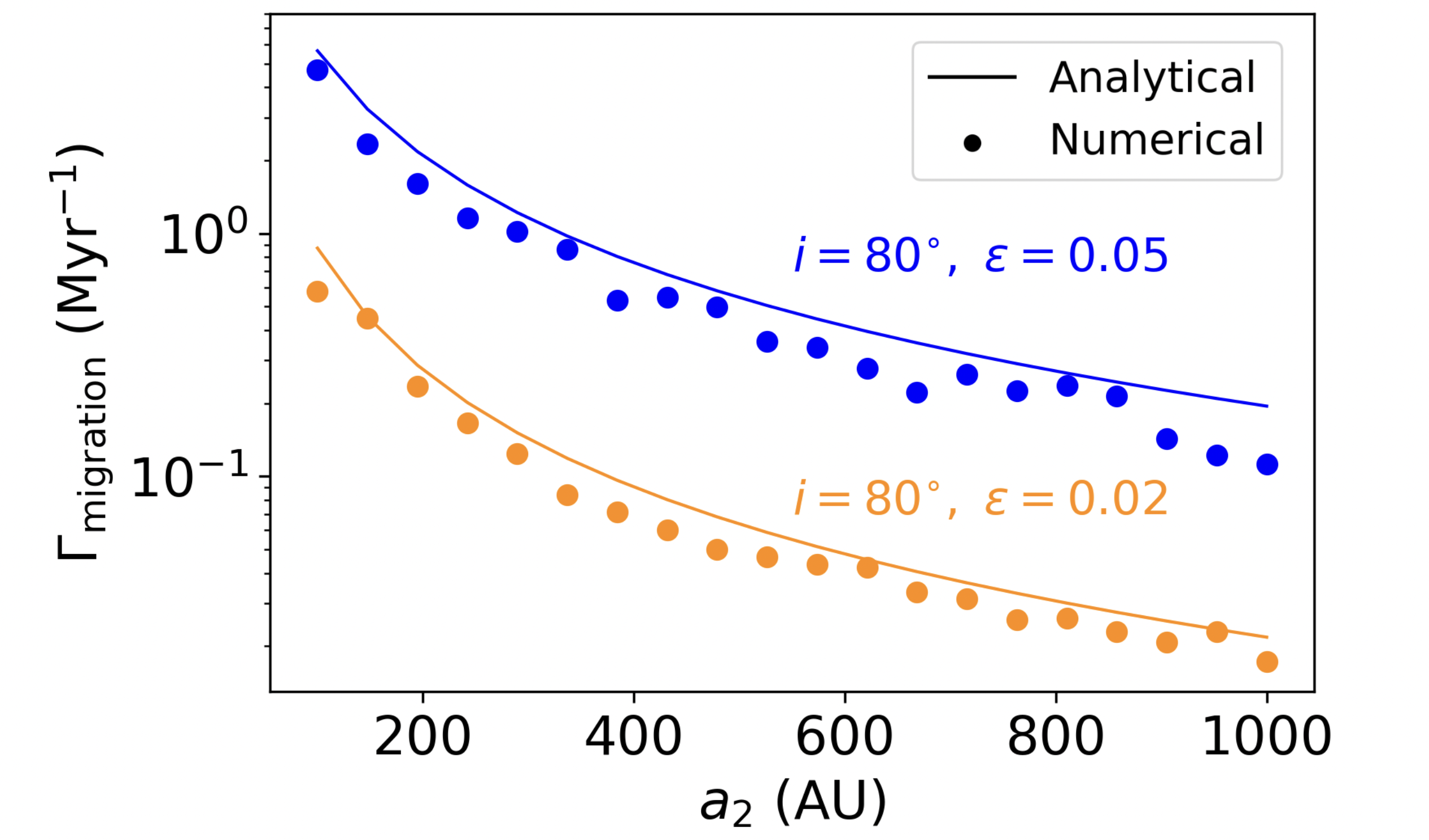}
\caption{\footnotesize Comparison of the numerical (dots) and analytical (solid curves) rates of migration for a consistent number of planetary systems (see \S \ref{sec:applications} for initial conditions). We see that the analytical prediction closely matches the numerical results. As $a_2$ increases, the rates decrease as expected due to the weaker influence of the stellar perturber.} 
\label{fig:rates}
\end{center}
\end{figure*}

We demonstrate that the model can be used to estimate the rate of Hot Jupiter migration for a population of systems, see Figure \ref{fig:rates}. We generate two ensembles of 500 systems each with initial conditions of $m_1 = 1M_{\odot}$, $m_2 = 1 M_J$, $m_3 = 1M_{\odot}$, $a_1 = 2$ AU, $a_2$ varied uniformly from 100 to 1000 AU, $e_1 = 0.01$, $i=80^{\circ}$, $\omega_1$ drawn uniformly from $0^{\circ}$ to $360^{\circ}$, and $\omega_2$ drawn uniformly from $0^{\circ}$ to $360^{\circ}$. For one set of systems, we fix $\epsilon = 0.02$ and for the other we fix $\epsilon = 0.05$. We choose these values because they lie in the regime sensitive to high-eccentricity excitation and produce different descent timescales. Each system is numerically evolved for 1000$t_{\rm quad}$.

We then divide each ensemble of systems into 20 bins with 25 systems each and numerically determine the rate of migration for each bin by taking $\Gamma_{\rm migration} = t_{\rm descent}^{-1}$ for the systems that undergo a descent, then normalize by the number of systems within the bin. The numerical rates are shown as the dots in Figure \ref{fig:rates}. To determine the rate of migration analytically, we take the reciprocal of Eq.~(\ref{eq:t_TP}), shown as the solid curves in Figure \ref{fig:rates}. As expected, the rates become lower as the stellar perturber becomes more distant and thus provides weaker perturbations. Importantly, the analytical model closely tracks the numerical results, with slight overestimation due to the fact that some of the numerically integrated systems begin in the chaotic regime before quickly becoming regular (as discussed in \S \ref{sec:TP}). Thus, these models provide an accurate method for calculating rates of migration and tidal disruption for populations of systems with varied initial conditions.

\section{Conclusion}
\label{sec:conclusion}

Triple body systems are prevalent in nature, from planetary to stellar to supermassive black hole scales. The EKL mechanism seems to largely describe the dynamics of these sets of systems \citep[e.g.,][]{Naoz16}. In this paper, we have presented analytical approximations for secular descents driven by the EKL mechanism in hierarchical three-body systems. These models can be used to estimate the secular descent of an object's pericenter and are applicable to a wide variety of astrophysical systems. 

EKL causes oscillations of eccentricity and inclination at both the quadrupole and octupole levels in the secular equations of motion. The octupolar envelope of the quadrupole oscillations describes the overall increase in the inner orbit's eccentricity, i.e., the overall decrease in the inner orbit's pericenter distance. For sufficiently high initial $i$ and $\epsilon$, the eccentricity can be driven to near unity during an octupole cycle.

We divide our discussions into two regimes, the test particle approximation, \S \ref{sec:testparticle}, of which $m_2\to 0$, and beyond the test particle, \S \ref{sec:beyondTP}. In the former approximation, we employed two approaches. The first approach, discussed in \S \ref{sec:TP_swing}, assumes small $J_z$ (near-perpendicular initial mutual inclination for an initially circular orbit) and provides an analytical description of the step-wise evolution of $J_z$. Using the evolution of $J_z$ in Eq.~(\ref{eq:Jz_swing}) and the corresponding step timescale in Eq.~(\ref{eq:t_step}), individual steps in $J_z$ can be stitched together to yield the overall secular descent, see Figure \ref{fig:Jz_model_example}. This approach is useful for understanding the increase in eccentricity during individual quadrupole cycles. The second test particle approach, discussed in \S \ref{sec:TP}, is valid for inclinations $\gtrsim 45^{\circ}$ and is based on integrating the equation of motions (see  Eq.~(\ref{eq:t_TP})). This approach yields the full secular descent of the orbit's pericenter. A numerical prefactor is obtained by calibrating to Monte Carlo simulations. We present a similar approach extended beyond the test particle case in \S \ref{sec:beyondTP}, and the analogous result is Eq.~(\ref{eq:t_nonTP}). As expected, the timescale for descent has a similar functional form between the test and non-test particle cases, with a characteristic octupole timescale
\begin{equation}
    t_{\rm oct}=\frac{64}{15} k^{-1}\frac{a_2^4}{a_1^{5/2}} \frac{(m_1+m_2)^{3/2}}{(m_1-m_2)m_3}  \frac{(1-e_2^2)^{5/2}}{e_2} \ ,
\end{equation}
which naturally comes from the derivation.

As depicted in Figure \ref{fig:examplesystem}, the two analytical approaches (for the test and non-test particle cases) agree well with numerical evolution for a representative system, and the analytical approach agrees with the numerical evolution to within a small factor over the entire parameter space sensitive to deep descents (see Figure \ref{fig:ei_grid}). While these approximations are valid for nearly regular (non-chaotic) trajectories, we perform numerical tests that show most initially chaotic systems tend to quickly become regular, and the approximations in this work accurately describe the regular portion of the trajectory (see Figure \ref{fig:omegagrid}).

In \S \ref{sec:applications}, we demonstrate the application of the model to Jupiter-like planets being secularly driven from large distances to the Roche limit by a stellar companion. In Figure \ref{fig:rates}, we show that the model can be used to accurately estimate rates of EKL-driven migration for a population of Hot Jupiters in stellar binaries. The ability to circumvent numerical simulations makes the models especially useful for understanding features of populations of systems. We note that the approximations can be applied to other astrophysical settings in which the EKL mechanism acts to reduce an object's pericenter distance, such as black hole systems that produce tidal disruption events or gravitational waves.

The authors gratefully acknowledge the anonymous referee for useful comments. We thank Barak Kol, Ygal Klein, Isabel Angelo, Thea Faridani, and Denyz Melchor for useful discussions. The authors thank the support of NASA XRP grant 80NSSC23K0262. S.N thanks Howard and Astrid Preston for their generous support. The authors also acknowledge the use of the UCLA cluster \textit{Hoffman2} for computational resources. This research has made use of NASA’s
Astrophysics Data System Bibliographic Services.

\newpage

\software{
    NumPy \citep{numpy},
    SciPy \citep{scipy},
    Matplotlib \citep{matplotlib},
    Mathematica \citep{Mathematica}
}

\bibliography{paperbib, softwarebib}{}
\bibliographystyle{aasjournal}

\appendix

\section{Equations of Motion}
\label{app:EOM}

Here we present the hierarchical three-body Hamiltonian beyond the test particle. The doubly-averaged Hamiltonian (averaged over both orbits) with the nodes eliminated ($\Omega_1-\Omega_2 = \pi$) up to octupole order is \citep[see e.g.,][]{Naoz16}
\begin{equation}
    \mathcal{H} = \mathcal{H_{\text{quad}}} + \epsilon_M \mathcal{H_{\text{oct}}} \ ,
\end{equation}
where the quadrupole component is 
\begin{equation}
    \mathcal{H}_{\text {quad }}=C_2\left\{\left(2+3 e_1^2\right)\left(3 \cos ^2 i-1\right)+15 e_1^2 \sin ^2 i \cos \left(2 \omega_1\right)\right\} \ ,
\end{equation}
the octupole component is 
\begin{equation}
    \begin{aligned} \mathcal{H}_{\text{oct}} & = \frac{15}{4} C_2 e_1 \left\{A \cos \phi+10 \cos i \sin ^2 i\left(1-e_1^2\right) \sin \omega_1 \sin \omega_2\right\} \ ,  \end{aligned}
\end{equation}
and
\begin{equation}
    \epsilon_M = \frac{m_1-m_2}{m_1+m_2} \frac{a_1}{a_2}\frac{e_2}{1-e_2^2} \ ,
\end{equation}
and 
\begin{equation} \label{eq:helpful_terms}
    \begin{aligned} L_1 & =\frac{m_1 m_2}{m_1+m_2} \sqrt{k^2\left(m_1+m_2\right) a_1} \\ L_2 & =\frac{m_3\left(m_1+m_2\right)}{m_1+m_2+m_3} \sqrt{k^2\left(m_1+m_2+m_3\right) a_2}  \\ G_1 & = L_1 \sqrt{1-e_1^2} \\ G_2 & = L_2 \sqrt{1-e_2^2} \\ C_2 & = \frac{k^4}{16} \frac{\left(m_1+m_2\right)^7}{\left(m_1+m_2+m_3\right)^3} \frac{m_3^7}{\left(m_1 m_2\right)^3} \frac{L_1^4}{L_2^3 G_2^3} \\ C_3 & =-\frac{15}{16} \frac{k^4}{4} \frac{\left(m_1+m_2\right)^9}{\left(m_1+m_2+m_3\right)^4} \frac{m_3^9\left(m_1-m_2\right)}{\left(m_1 m_2\right)^5} \frac{L_1^6}{L_2^3 G_2^5} \\ A & =4+3 e_1^2-\frac{5}{2} B \sin i^2 \\ B & = 2+5e_1^2-7e_1^2 \cos 2\omega_1 \\ \cos \phi & =-\cos \omega_1 \cos \omega_2-\cos i \sin \omega_1 \sin \omega_2 \ .
    \end{aligned}
\end{equation}
The full equations of motion can be found in \cite{Naoz16}. In \S \ref{sec:beyondTP}, we use the eccentricity evolution of the inner orbit given by
\begin{equation}
\begin{aligned}
\frac{de_1}{dt} & =C_2 \frac{1-e_1^2}{G_1}\left[30 e_1 \sin ^2 i \sin \left(2 \omega_1\right)\right] \\
& +C_3 e_2 \frac{1-e_1^2}{G_1}\left[35 \cos \phi \sin ^2 i e_1^2 \sin \left(2 \omega_1\right)\right. \\
& -10 \cos i \sin ^2 i \cos \omega_1 \sin \omega_2\left(1-e_1^2\right) \\
& \left.-A\left(\sin \omega_1 \cos \omega_2-\cos i \cos \omega_1 \sin \omega_2\right)\right] \ .
\label{eq:dedt_full}
\end{aligned}
\end{equation}

Moving to the test particle limit ($m_2 \to 0$), the test particle Hamiltonian is defined in \S \ref{sec:picture}. Using the formalism presented there, where $J=\sqrt{1-e_1^2}$ and $J_z = \theta \sqrt{1-e_1^2}$, the equations of motion for a test particle are 
\begin{eqnarray}
    \frac{d J}{d t}&=&\frac{\partial F}{\partial \omega_1} \label{eq:dJdt_TP}  \\ \frac{d J_z}{d t}&=&\frac{\partial F}{\partial \Omega_1} \label{eq:dJzdt_TP} \\ \frac{d \omega_1}{d t}&=&\frac{\partial F}{\partial e_1} \frac{J}{e_1}+\frac{\partial F}{\partial \theta} \frac{\theta}{J} \label{eq:dom1dt_TP} \\ \frac{d \Omega_1}{d t}&=&-\frac{\partial F}{\partial \theta} \frac{1}{J} \ .
\end{eqnarray}
The conversion from $t$ to true time $T$ is given by 
\begin{equation} \label{eq:truetime}
    t \equiv T \frac{3}{8} \frac{m_3}{m_1} \Omega_* \left(\frac{a_1}{a_2}\right)^3 \frac{1}{(1-e_2^2)^{3/2}} \ ,
\end{equation}
where $\Omega_*$ is the orbital angular speed of the test particle.

\section{Fitting the model}
\label{app:fitting}

\begin{figure*}
\begin{center}

\includegraphics[width=7in]{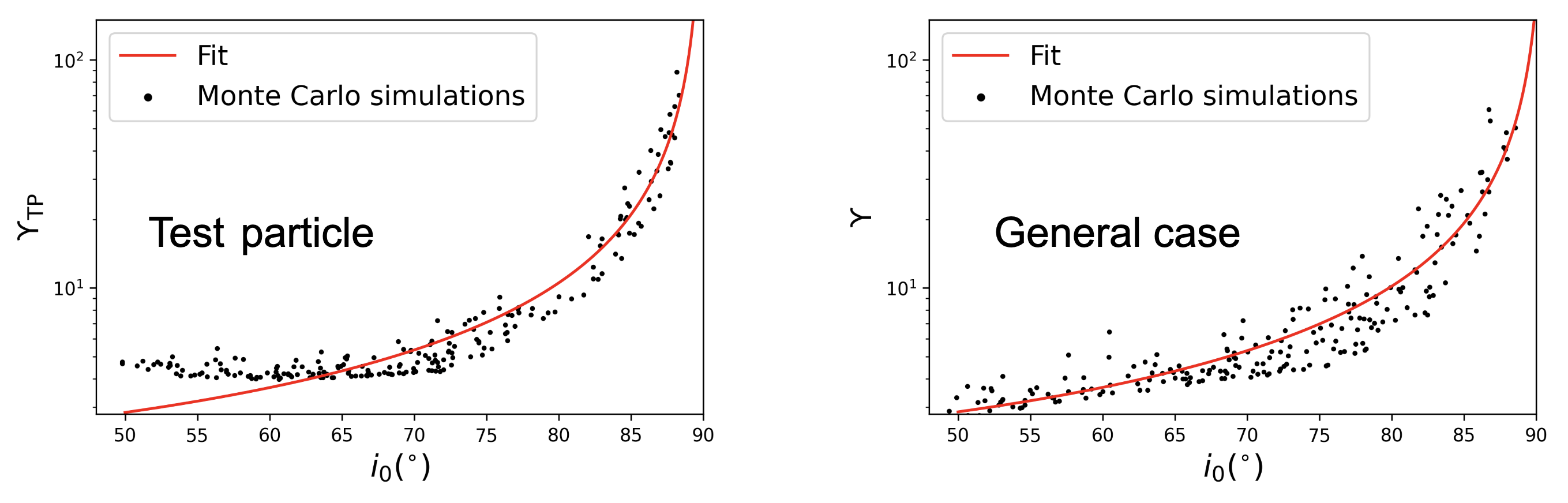}
\caption{\footnotesize Fits to $\Upsilon$ obtained from Monte Carlo simulations in the test particle case (left) and general case (right). The procedure is described in Appendix \ref{app:fitting}. Fitted values of $\Upsilon$ can be used in e.g., Eq.~{\ref{eq:integral_TP}} to estimate the secular descent. The average error in each case is $\sim$25\%, with the test particle fit diverging to a factor of $\sim$2-3 below $\sim$60$^{\circ}$. For the general case fit displayed here, we use a fiducial value of $G_1/G_2 = 0.01$, but the exact value was calculated for each system to obtain the fit. In these simulations, systems below $\sim$50$^{\circ}$ do not flip as they require a value of $\epsilon$ that exceeds the stability criterion to induce a flip.}
\label{fig:fitting}
\end{center}
\end{figure*}

We fit the models in \S \ref{sec:TP} and \S \ref{sec:beyondTP} by comparing to Monte Carlo simulations. For the test particle case, we generate a population of 500 systems uniformly drawn with initial conditions of $m_1 = 1 M_{\odot}$, $m_3 = 0.1 - 1.0 M_{\odot}$, $a_1 = 1$ AU, $a_2 = 8-25$ AU, $e_1 = 0.01-0.1$, $e_2 = 0.1 - 1$, and $i = 45-90^{\circ}$. These parameters are chosen to represent a range of values in which deep descents are able to occur. We repeat for the general case and draw $m_2$ log-uniformly from $10^{-3}-10^{-0.5} M_{\odot}$. When sampling, we enforce the stability criterion of \cite{Mardling+01}:
\begin{equation}
    \frac{a_2}{a_1}>2.8\left(1+\frac{m_3}{m_1+m_2}\right)^{2 / 5} \frac{\left(1+e_2\right)^{2 / 5}}{\left(1-e_2\right)^{6 / 5}}\left(1-\frac{0.3 i}{180^{\circ}}\right) \ .
\end{equation}
For the systems that undergo a flip on the initial descent (and thus display large eccentricity evolution that can be robustly fitted), we estimate $\Upsilon$ by comparing the timescale to reach the maximum eccentricity at the flip in the simulations with the timescale in Eq.~(\ref{eq:t_TP}) (test particle) and Eq.~(\ref{eq:t_nonTP}) (general case) to reach that eccentricity. We find that $\Upsilon$ depends most sensitively on the initial mutual inclination $i_0$ and is well-characterized by the inclination modification term for the octupole timescale from \cite{Teyssandier+13}:
\begin{equation}
    \Upsilon = \frac{A}{\frac{G_1}{G_2}+\cos{i_0}},
\end{equation}
where $A$ is a numerical coefficient, and $G_1$ and $G_2$ are defined in Eq.~(\ref{eq:helpful_terms}). This dependence appears in the ratio between the quadrupole and octupole terms in the equations of motion for the arguments of periastron \citep[e.g.,][]{Naoz16}, and is thus a natural choice for scaling the octupole timescale in this work. For the test particle case, we obtain $A = 1.83$, In the test particle case, $G_1 = 0$, so $\Upsilon$ only depends on $i_0$. For the general case, we calculate $G_1/G_2$ for each system and obtain a numerical factor of $A = 1.87$. The fits to the simulations are shown in Figure \ref{fig:fitting}. We find that the average error in both cases is $\sim$25\%, with the test particle fit diverging to a factor of $\sim$2-3 below $\sim$60$^{\circ}$, and the general fit being roughly correct at all inclinations between 50-90$^{\circ}$. 

\section{Other physical timescales}
\label{app:timescales}

\begin{figure*}
\begin{center}

\includegraphics[width=7in]{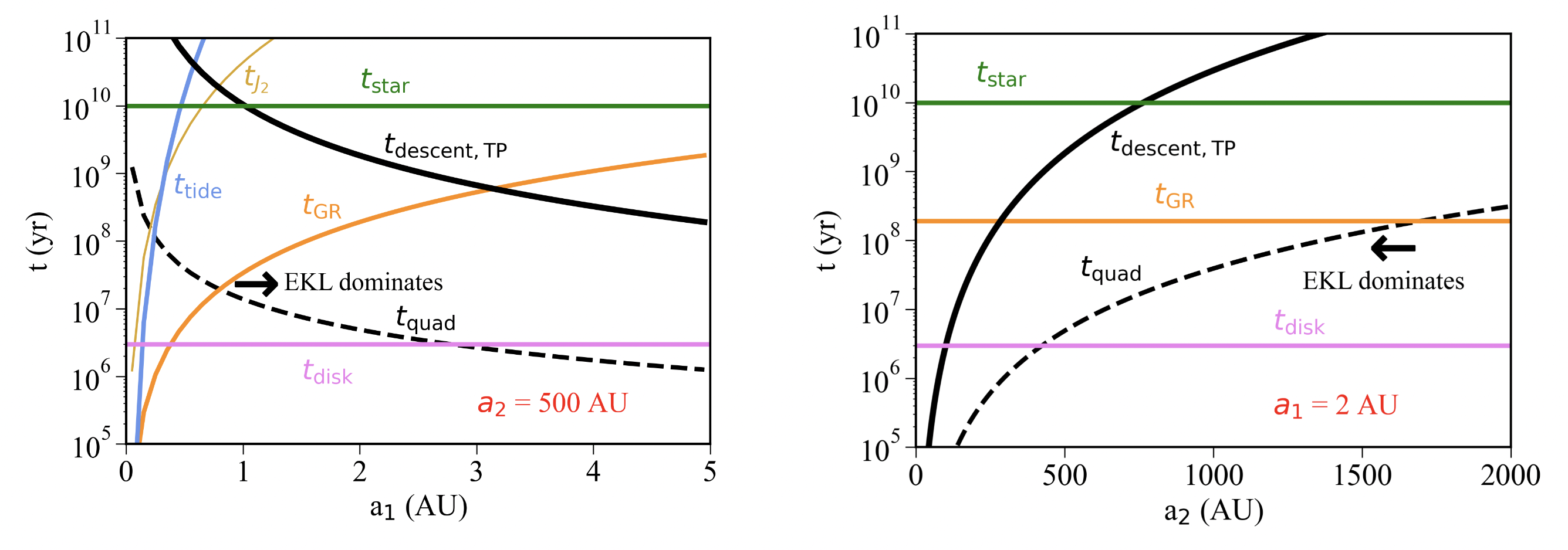}
\caption{\footnotesize Example calculations of of the time to descend to the Roche limit, $t_{\rm descent,TP}$, from Eq.~(\ref{eq:t_TP}) (solid black curve). We compare with other relevant physical timescales for a Jupiter-like planet ($m_2 = 1 M_J$) in a stellar binary ($m_1=m_3 = 1M_{\odot}$). In both panels, we set $e_1 = 0.01$, $e_2 = 0.5$, and $i=70^{\circ}$. In the left panel, we fix $a_2 = 500$ AU and vary $a_1$. In the right panel, we fix $a_1 = 2$ AU and vary $a_2$. We compare with the quadrupole timescale from Eq.~(\ref{eq:t_quad}) (dashed black curve), the GR precession timescale from Eq.~(\ref{eq:t_GR}) (orange curve), the tidal precession timescale from Eq.~(\ref{eq:t_tide}) (blue curve), the $J_2$ precession timescale from Eq.~(\ref{eq:t_J2}) (yellow curve), a typical disk dissipation timescale of 3 Myr (violet curve), and the main-sequence lifetime of a Sun-like star (green curve). GR dominates when $t_{\rm GR}<t_{\rm quad}$, and tides and $J_2$ become important at low $a_1$. The descent timescale is comparable to the disk lifetime when $a_1/a_2$ is small. For most systems, the Jupiter arrives at a close distance after the disk dissipates but during the star's lifetime. Here, we see that the model provides an efficient method for estimating features of the Hot Jupiter population, such as the age of Hot Jupiters post-migration and the system configurations in which high-eccentricity migration is expected to occur.} 
\label{fig:timescales}
\end{center}
\end{figure*}

In Figure \ref{fig:timescales}, we compare the secular descent timescale (solid black line) with other physical timescales. Specifically, general relativity (GR) causes the inner orbit to precess in the opposite direction of the EKL precession, potentially suppressing the EKL eccentricity excitations. The GR precession timescale of the inner orbit at the first Post-Newtonian level can be estimated as \citep[e.g.,][]{Naoz+13b}
\begin{equation} \label{eq:t_GR}
    t_{\rm GR} \approx 2 \pi \frac{a_1^{5/2}c^2(1-e_1^2)}{3k^3(m_1+m_2)^{3/2}} \ ,
\end{equation}
where $c$ is the speed of light. It was noted before, e.g., \citet{Naoz+13b} and \citet{Hansen+20}, that EKL eccentricity excitations are suppressed roughly when $t_{\rm GR}<t_{\rm quad}$, where $t_{\rm quad}$ is the quadrupole timescale from Eq.~(\ref{eq:t_quad}). This timescale is shown in Figure \ref{fig:timescales} as a light orange line. 

Another physical process that affects the evolution is tidal precession, circulation, and shrinking of the semi-major axis. The tidal precession can suppress EKL eccentricity excitations. The tidal precession timescale (shown as a blue line in Figure \ref{fig:timescales}) is estimated as \citep[e.g.,][]{Eggleton+98,Naoz+14}
\begin{equation} \label{eq:t_tide}
    t_{\rm tide} \approx \frac{a_1^{13/2}(1-e_1^2)^{5}m_1m_2}{15k\sqrt{m_1+m_2}\left(1+\frac{3}{2}e_1^2 + \frac{1}{8}e_1^4\right)\Lambda} \ ,
\end{equation}
where
\begin{equation}
    \Lambda=m_2^2k_{L,1}R_1^5 + m_1^2 k_{L,2}R_2^5 \ ,
\end{equation}
where $R_1$ is the radius of the primary body and $k_{L,1}$ ($k_{L,2}$) is the apsidal motion constant of the primary (secondary) body. Tides become important when $t_{\rm tide}$ is comparable to the other dynamical timescales, which occurs for very low $a_1$. In the high-eccentricity migration formation channel for Hot Jupiters, tidal forces can shrink and circularize the inner orbit.

Additionally, the oblate potential of the star can cause the planetary orbit to precess. This timescale (shown as a yellow line in Figure \ref{fig:timescales}) is estimated as \citep[e.g.,][]{Murray+99}
\begin{equation} \label{eq:t_J2}
    t_{J_2} \approx \frac{4 \pi}{3} \frac{1}{nJ_2} \left( \frac{a_1}{R_1} \right)^2 ,
\end{equation}
where $n$ is the mean motion of the planet and $J_2$ is the gravitational quadrupole moment of the primary body. For the solar value of $J_2 \approx 2 \times 10^{-7}$ \citep[e.g.,][]{Godier+99}, the $J_2$ precession becomes relevant only for very low $a_1$ \citep[for the consequences of evolving $J_2$, see e.g.,][]{Faridani+23}.

We also compare to relevant lifetimes within a Hot Jupiter system. A major question concerns the role of a gas disk in forming Hot Jupiters, either as a distinct formation channel \citep[e.g.,][]{Lin+86} or as an aid to the high-eccentricity migration process \citep[e.g.,][]{Vick+23}. The typical timescale for a gas disk to dissipate is $\sim$3 Myr \citep[e.g.,][]{Fedele+10}, shown as a magenta line in Figure \ref{fig:timescales}. We find that most Hot Jupiters formed via EKL are delivered to the Roche limit after the disk dissipates. We also compare to the $\sim$10 Gyr lifetime of a Sun-like main sequence star (green line in Figure \ref{fig:timescales}) and see that Hot Jupiters migrate over a large range of timescales depending on the system configuration.

\end{document}